# Machine Learning Recognition of hybrid lead halide perovskites and perovskite-related structures out of X-ray diffraction patterns

Marchenko E.I.[a,b], Korolev V.V.[c], Kobeleva E. A.[b], Belich N.A. [a], Udalova N.N. [a], Eremin N.N. [b,e], Goodilin E.A.[a,d], Tarasov A.B.[a,d*]

[a] *Laboratory of New Materials for Solar Energetics, Department of Materials Science, Lomonosov Moscow State University; 1 Lenin Hills, 119991, Moscow, Russia;*
[b] *Department of Geology, Lomonosov Moscow State University; 1 Lenin Hills, 119991, Moscow, Russia*
[c] *MSU Institute for Artificial Intelligence, Lomonosov Moscow State University; 119192, Moscow, Russia*
[d] *Department of Chemistry, Lomonosov Moscow State University; 1 Lenin Hills, 119991, Moscow, Russia*
[e] *Institute of Geology of Ore Deposits, Petrography, Mineralogy, and Geochemistry, Russian Academy of Science, Moscow, Russia*
e-mail: alexey.bor.tarasov@yandex.ru

**Abstract**

Identification of crystal structures is a crucial stage in the exploration of novel functional materials. This procedure is usually time-consuming and can be false-positive or false-negative. This necessitates a significant level of expert proficiency in the field of crystallography and, especially, requires deep experience in perovskite - related structures of hybrid perovskites. Our work is devoted to the machine learning classification of structure types of hybrid lead halides based on available X-ray diffraction data. Here, we proposed a simple approach to quickly identify of dimensionality of inorganic substructures, types of lead halide polyhedra connectivity and structure types using common powder XRD data and ML - decision tree classification model. The average accuracy of our ML algorithm in predicting the dimensionality of inorganic substructure, type of connection of lead halide and inorganic substructure topology by theoretically calculated XRD pattern among 14 most common structure types reaches 0.86±0.05, 0.827±0.028 and 0.71±0.05, respectively. The validation of our decision tree classification ML model on experimental XRD data shows the accuracies of 1.0 and 0.82 for the dimension and structure type prediction. Thus, our approach can significantly simplify and accelerate the interpretation of highly complicated XRD data for hybrid lead halides.

**Introduction**

The process of determining of crystal structures through X-ray diffraction (XRD) involves a series of intricate steps, including indexing, space group determination, structure solving, and structural model refinement. This complex procedure is time-consuming and requires certain expertise in crystallography. Often, the most difficult step is refining the structure using the Rietveld method [1]. The conventional approach to crystal structure determination via X-ray diffraction necessitates the manual optimization of numerous parameters. Additionally, the determination of the space group during the initial phase of structural analysis commonly involves manual trial-and-error procedures. Given that, these time-consuming processes a bottleneck hindering the rapid discoveries of new materials [2]. Minimizing human involvement in these procedures COULD enhance ITS efficiency and facilitate the implementation of high-throughput experimental approaches. Application of ML for diffraction data analysis is a hot research topic in recent times [3–6]. Recent progress in machine-learning (ML) techniques also have introduced innovative approaches to predict symmetry or unit cell parameters of crystal structures by XRD patterns data, chemical formulas and other crystal chemical descriptors [5–13]. Nevertheless, there remains a notable gap in the literature regarding the efficient identification of structural types for unknown materials solely from their powder X-ray diffraction patterns. The development of such a method could revolutionize high-throughput strategies for material discovery, opening up a new avenue for accelerated research.

In this work, we for the first time propose a machine learning methods for the automatic interpretation of powder diffraction data for such a class of compounds as hybrid lead halides. A group of such compounds called as hybrid perovskites and perovskite-related phases have garnered considerable interest as a prospective materials in the field of optoelectronics and photovoltaics due to their notable structural diversity, leading to increased research focus over the past decade [14–17]. These materials are composed of a combination of organic and inorganic components, typically consisting of lead, halide ions (such as chloride, bromide, or iodide), and organic cations (such as methylammonium or formamidinium). The unique properties of hybrid lead halide perovskites, such as their high absorption coefficients, long carrier diffusion lengths, and tunable bandgaps, make them promising candidates for use in solar cell devices [18–21].

The synthesis of hybrid metal halides represents a relatively uncomplicated procedure that is amenable to high-throughput screening [22, 23]. However, an effective tool for rapid automatic identification of the structure type of lead hybrid halides from powder XRD patterns has not previously been developed. This is because this class of compounds exhibits a wide variety of inorganic sublattice structural motifs, making it difficult to select descriptors for a ML model training. Despite this, in [24], the authors proposed to classify all hybrid halide structures into "perovskite" and "non-perovskite" groups from XRD data using ML based models (random forest and convolutional neural network). Such a classification is often used in perovskite community, however, it does not provide comprehensive information regarding the crystal structure and structure-property relationships. It is worth noting that to divide all crystal structures of hybrid lead halides into only two groups [25]: perovskite (built from only corner-shared lead halide octahedra) and non-perovskite (compounds with edge-sharing, face-sharing, or isolated lead halide octahedra) is not correct from a crystallographic point of view.

In this study, a machine learning algorithm of hybrid perovskite structures classification was developed and trained to predict the dimensionality of inorganic substructure motifs, connectivity type of lead halide octahedra, and structure types based on their calculated XRD patterns using a constructed dataset if hybrid lead halides with different structure types. The accuracy of the model's predictions was validated using experimental diffraction data. Thus, our approach shows that for a certain class of compounds, such as hybrid lead halides, structure types (structural motifs) can be automatically determine by ML classification models from the XRD patterns.

**Results and Discussion**

*Framework for XRD classification*

Our schematic framework for classification of XRD patterns of hybrid lead halides according to three structural descriptors (dimensionality of inorganic substructure, lead halide octahedra connectivity type and structure type) is shown in Figure 1. The algorithm of classification within the framework is divided into 8 steps: 1) dataset construction, 2) simulation of XRD powder patterns, 3) data filtering by significant XRD peaks, 4) obtaining of experimental XRD spectra, 5) crystal structure refinement, 6) manual labeling of experimental XRD spectra, 7-8) training and testing the machine learning model for crystal structure descriptors prediction. The methodology employs a combination of experimental and simulated XRD patterns to test the ML classification algorithm.

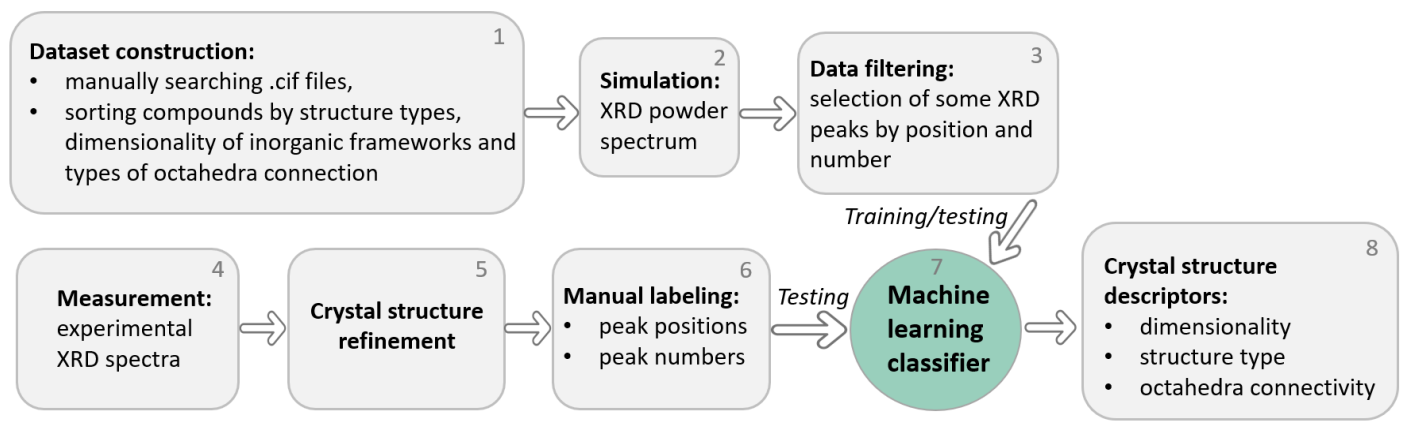

Figure 1. Schematic framework of our XRD data classification algorithm.

*Dataset construction*

In total, the constructed dataset contains 272 representative hybrid lead halide crystal structures [20, 24, 26–29]. Crystal structures with iodine dominate in the database compared to other halogens. To train and test the ML model, we analyzed the dataset according to three main characteristics of inorganic sublattices: 1) dimensionality, 2) lead halide octahedra connection types, 3) topology (structure type). The accuracy of this categorization was verified through a manual validation. Among all structures in the dataset, 14 ones belong to 0D, 52 - 1D, 141 - 2D and 65 - 3D dimensionality of inorganic sublattice (Figure 2a). There are 6 different groups of structures with different lead halide octahedra connectivity: isolated, connected by faces, vertices, vertices and faces, edges, edges and faces. Upon examination of the type of connection of lead halide octahedra in the dataset, it was observed that the distribution of data is not uniform, with the predominant type of connection by vertices (Figure 2b).

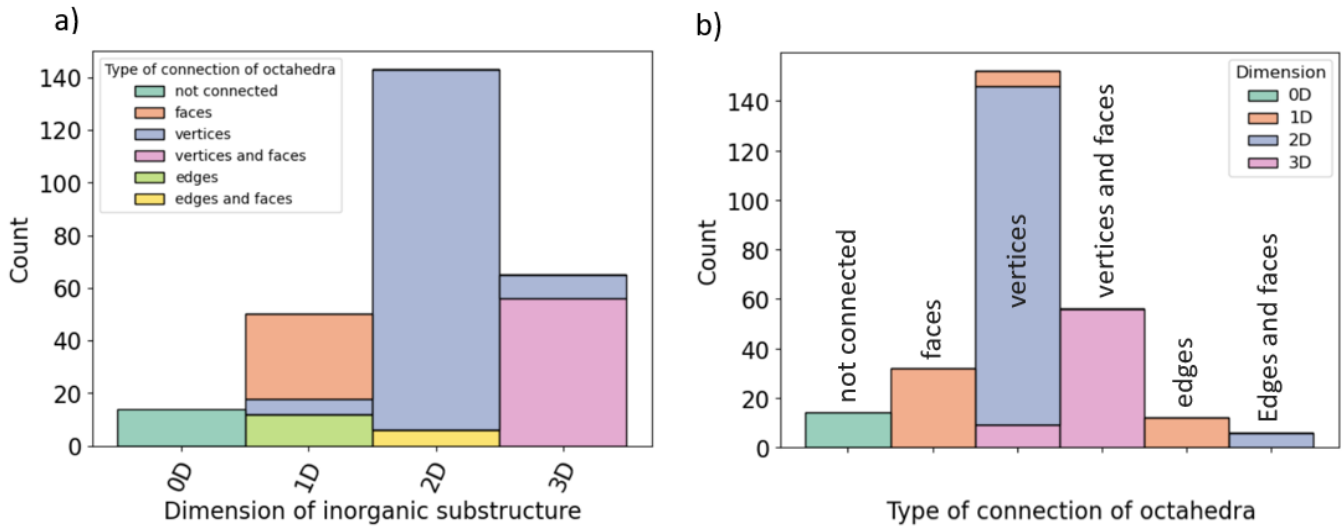

Figure 2. Distribution of data in dimensionality of inorganic substructures (a) and types of connection of octahedra (b).

At the first step, we compiled the dataset of 14 different most common structure types from the following resources: the Database of 2D hybrid halide perovskites of the laboratory of new materials for solar energetics (NMSE) (24 samples of (100) with n=1, 34 samples of (100) with n=2, 30 samples of (100) with n=3, 11 samples of (100) with n=4, 30 samples of (110) 2X2 and 6 samples of (110) 3X3), 56 samples of experimentally known and theoretically predicted stoichiometric to perovskite $APbX_3$ polytypes according to [27], 9 samples of 3D perovskite structure type [29], 6 examples of 2D layered structures with connection of octahedra by edges and faces [28], 14 - 0D structures with isolated octahedra , 33 – 1D structures with chains of octahedra connected along faces, 7 – 1D double structures with chains of octahedra connected along vertex and 5 1D structures with chains of octahedra connected along edges [20, 28] (Table 1, Figure 3). The range of structure types presented in the dataset is not exhaustive, as there are approximately 30 distinct structure types of hybrid lead halides. A structure type was incorporated into the dataset and used for ML model training and testing when a group of a given structure type of a minimum of 5 crystal structures featuring diverse large cations is found. Our dataset can further be expanded with the emergence of newly experimentally refined structure types of hybrid lead halides.
This number of structures in the dataset was enough to train and test our ML model.

Table 1. Distribution of data on dimensionality, type of connection of octahedra and topological structure types.

| № | Dimensionality of inorganic sublattice | A number of structures in dataset | Topology of inorganic sublattice or structure type |
|---|---|---|---|
| 1 | 0D | 14 | isolated octahedra |
| 2 | 1D | 33 | chains of octahedra connected along faces |
| 3 | 1D | 7 | chains of octahedra connected along vertices |
| 4 | 1D | 5 | chains of octahedra connected along edges |
| 5 | 1D | 7 | double chains of octahedra connected along edges |
| 6 | 2D | 30 | (100) type with n=3 |
| 7 | 2D | 34 | (100) type with n=2 |
| 8 | 2D | 24 | (100) type with n=1 |
| 9 | 2D | 30 | (110) corrugated 2X2 |
| 10 | 2D | 6 | layers of octahedra connected along vertices and faces |
| 11 | 2D | 6 | (110) corrugated 3X3 |
| 12 | 2D | 11 | (100) type with n=4 |
| 13 | 3D | 56 | hexagonal close packed with more than 3 layers in close packaging |
| 14 | 3D | 9 | perovskite structure type |

It is important to stress, that the crystal structures with a high disorder of organic cations and partially occupied atomic sites are also suitable for training ML algorithm, since we use the representation of the crystallographic data in the form of a "fingerprint" – simulated XRD diffraction pattern.

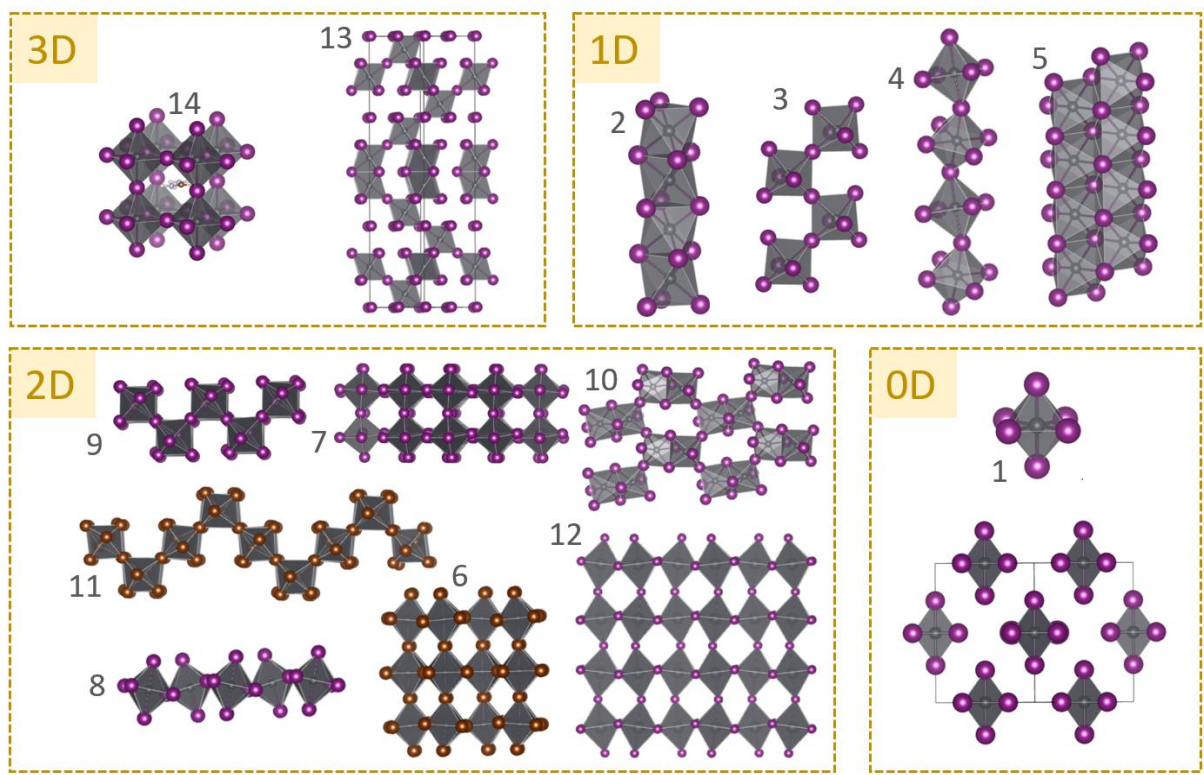

Figure 3. Different most common inorganic substructure motifs of hybrid lead halides contain in the dataset. Numbers of structure types according to Table 1 are indicated in gray. Organic cations were removed for clarity.

*Signification of the important XRD features and ML model construction*

An automatically selection of XRD patterns features for the ML model training are not always reasonable. Therefore, to identify the most important features of hybrid lead halide XRD patterns, we selected the features manually. The most representative peaks for all the structures from the dataset are located in the range of up to 30-40° 2θ and contain a few reflections of structures with a high symmetry or up to 200 reflections for triclinic and monoclinic structures (Figure S1). The powder X-ray diffraction pattern for each compound of this database was simulated from 3.0 ° to 30 ° 2θ using Pymatgen XRD calculator [30] considering a Cu Kα radiation. In this work, we used as key features the first 10 reflexes from simulated XRD pattern for dataset structures with a relative intensity greater than 0.05 in a.u. It is important to note, that peak intensities were not taken into account when training ML model. This decision was reached based on the observation that the texturing of crystallites in empirical X-ray diffraction (XRD) data often leads to significant deviations in reflection intensities when compared to theoretical (simulated) reference values. The symmetry group at this stage of the ML model also was not selected as an important feature, since the data among the symmetry groups is distributed extremely unevenly. Moreover, the symmetry of the inorganic substructure often differs from the symmetry of the structure for a given class of compounds [31].

In this work we used a decision tree (DT) classifier from the Scikit-learn (Python library) [32, 33] as a supervised classification ML algorithm. The training and testing datasets represented 80 % and 20 % of the data, respectively. A fivefold cross validation coupled approach was used to optimized hyperparameters and model.

*Dimensionality and octahedra connectivity prediction*

The developed DT model showed an accuracy (measures the proportion of correctly classified instances, providing an overall assessment of the algorithm's performance) of 0.76±0.07 (Figure 4a) if discriminating inorganic sublattice dimensionality into four classes (0D, 1D, 2D and 3D). It should be noted, that considering structures without 0D inorganic substructures, the average accuracy of ML model is 0.86±0.05 (Figure 4b). This outcome can be attributed to the prevalence of inorganic hybrid halides within the sampled 1D structures.

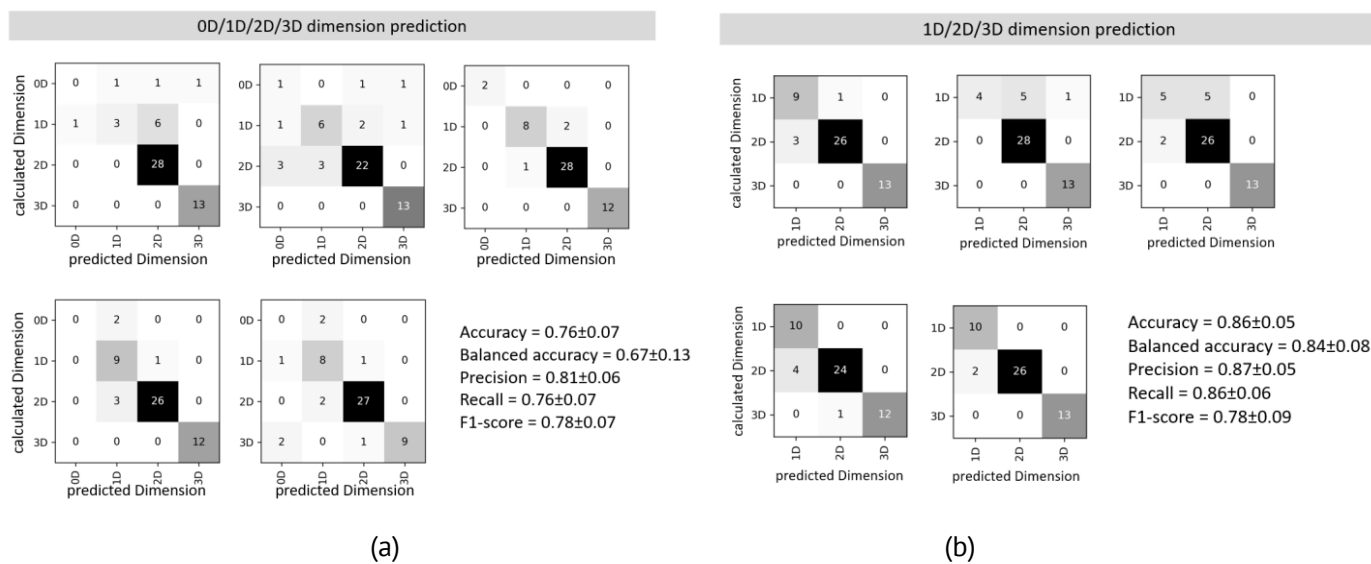

(a)     (b)

Figure 4. DT confusion matrix of inorganic sublattice dimension prediction of hybrid lead halides by XRD data using DT classification ML algorithm and fivefold cross validation coupled: classification by 0D, 1D, 2D and 3D (a), classification by 1D, 2D and 3D (b).

It is worth noting that evaluating the performance of a machine learning model involves considering multiple factors. The accuracy metric of the overall correctness of predictions may not adequately reflect the DT model performance in cases where class distribution is imbalanced. Therefore, other quality metrics applicable to our ML classification model should be provided. All values of our model quality metrics (accuracy, balanced accuracy, precision, recall, F1 score and MCC) are in Table S2. Thus, the quality metrics of DT model classifying dimensions of inorganic substructures into 4 classes are as follows: accuracy = 0.76±0.07, balanced accuracy = 0.67±0.13, precision = 0.81±0.06, recall = 0.76±0.07, F1-score = 0.78±0.07. However, when train and test DT model using three classes (1D, 2D and 3D), the values of all model quality metrics improved (Figure 4b). Thus, our DT model correctly classifies the dimension of the inorganic substructure among hybrid lead halides compounds.

Our ML algorithm also predicts the types of connections of lead halide octahedra by XRD pattern with high accuracy. The average accuracy and recall of predicting the type of connection of lead halide octahedra is 0.827±0.028 (Figure 5).

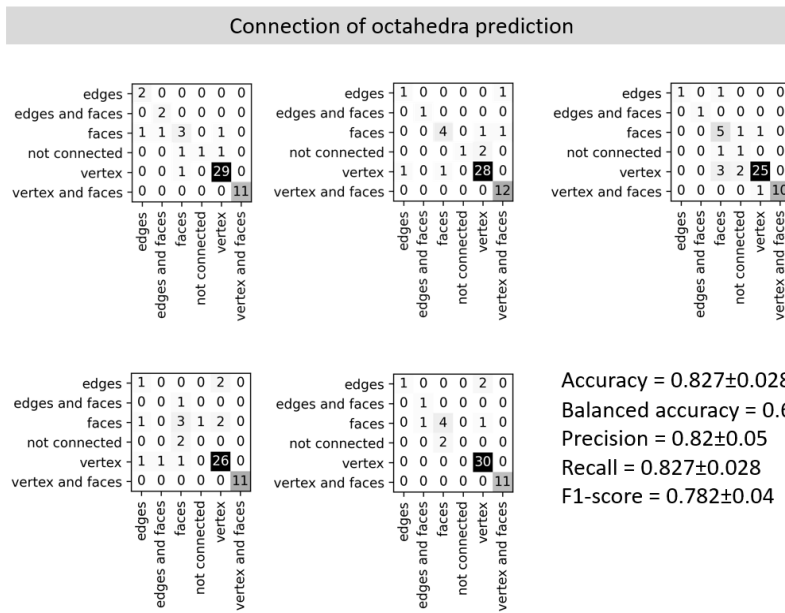

Figure 5. DT confusion matrix of prediction of type of lead halide octahedra connection prediction by XRD data using DT classification ML algorithm and fivefold cross validation coupled.

*Inorganic substructure topology prediction*

In our work, we used in the first time the ML classification model to classify the topology of inorganic substructures of hybrid lead halides. Here we trained and tested the DT model on 14 different structure types by inorganic sublattice topology. Our DT model showed an average accuracy of 0.71±0.05 and best accuracy of 0.8 (Figure 6a) at discriminating inorganic substructure topology according to Table 1. This outcome is attributed to the limited availability of crystal structures for specific structural categories. Enhancements of the algorithm could be achieved in subsequent iterations by augmenting the dataset with additional experimentally elucidated linkages.

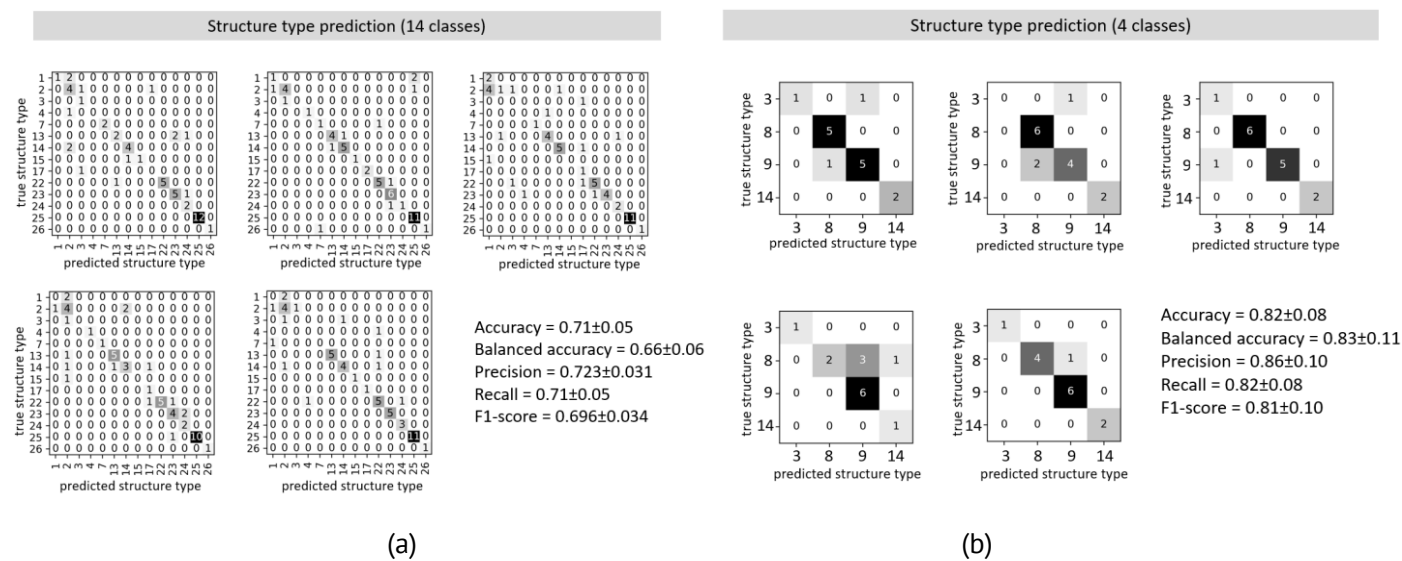

Figure 6. Confusion matrix of prediction of topology of inorganic sublattice structural motifs: training and testing model on 14 structure types (a) and the result for 4 structure types model testing (b).

Nevertheless, when the training/testing set size is constrained by specific structural types and the DT model is trained on the four predominant structural types with more than five structures each, alongside a test sample comprising more than two crystal structures of identical topological types, the algorithm demonstrates satisfactory accurate performance (Figure 6b). In this case the DT model showed an average accuracy of 0.82±0.08 (Figure 6b) at discriminating inorganic substructure topology among the four structure types that dominate the dataset.

Consequently, it can be concluded that the methodology proposed in this study is correct, as the DT algorithm exhibits sufficiently stable functionality. The accuracy of this approach can further increase by the augmentation the dataset with new emerging structural data from scientific literature sources.

*Validation DT algorithm on experimental data*

In addition to the tests on 272 simulated XRD patterns, the ML approaches were further validated on experimental XRD data. For the purpose of validation and to assess the generalization capabilities of the DT model, an additional set of 11 XRD patterns from experimental powder and thin film samples was used. These XRD patterns were not included in the initial training or testing sets of the model. 7 experimental diffraction powder patterns with 3D, 2D and 1D inorganic substructure topology obtained were taken from [24] and 4 XRD patterns (figure S2) from thin films were obtained in this work (compound 1 - $BA_2PbI_4$, compound 2 - $(F-PMA)_2PbI_4$, compound 3 - $GUACsPbI_4$ and compound 4 - $MAPbI_3$). The crystal structures of these compounds belong to the 3 most common structure types discussed earlier: 2D inorganic substructure of (100) type with n=1, 1D inorganic substructure with chains of octahedra connected along vertices and 3D crystal structures with perovskite structure type (Table 2, Figure S3). It is worth noting that due to the strong texturing of the experimental samples and the relatively low quality of the diffraction patterns data, we did not expect high accuracy of our ML algorithm, because it was training and testing on the ideal theoretical diffraction patterns. Surprisingly, the decision tree classification model reached accuracies of 1.00 and 0.82 for dimension and structure type prediction (figure 7a, b), respectively, for those 10 experimental XRD of synthesized samples Therefore, the observed accuracies persist at a relatively high level despite potential deteriorating factors such as preferential orientation, texturing and variations in signal-to-noise ratio on XRD patterns that may have influenced the classification of experimental data.

Table 2. Experimental samples data used for decision tree algorithm validation.

| Composition | Dimensionality of inorganic substructure | Topology of inorganic sublattice or structure type | Reference |
|---|---|---|---|
| MBA-16sub8 | 2D | 8. (100) type with n=1 | [24] |
| MBA-25sub9 | 2D | 8. (100) type with n=1 | [24] |
| RG134_E6Cl | 1D | 3. chains of octahedra connected along vertices | [24] |
| RG135_E6Cl | 1D | 3. chains of octahedra connected along vertices | [24] |
| RG137_E6Cl | 3D | 14. perovskite structure type | [24] |
| SK-MAPbCl | 3D | 14. perovskite structure type | [24] |
| TB011 | 2D | 8. (100) type with n=1 | [24] |
| $BA_2PbI_4$ (compound 1) | 2D | 8. (100) type with n=1 | This study |
| $(F-PMA)_2PbI_4$ (compound 2) | 2D | 8. (100) type with n=1 | This study |
| $GUACsPbI_4$ (compound 3) | 2D | 8. (100) type with n=1 | This study |
| $MAPbI_3$ (compound 4) | 3D | 14. perovskite structure type | This study |

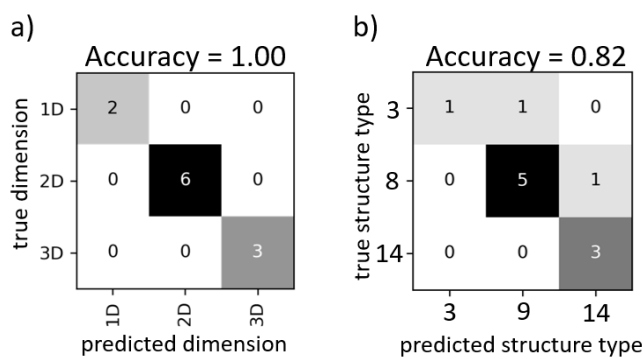

Figure 7. Confusion matrix for inorganic substructure dimension (a) and structure type (b) prediction by decision tree algorithm.

*Synthesis and experimental XRD patterns from thin films*

To evaluate the ML performances on experimental XRD powder patterns, 11 samples of synthesized hybrid lead halides were used. For 7 experimental samples data was taken from [24] and 4 samples were synthesized manually: compound 1 - $BA_2PbI_4$ ($BA^+$ – butylammonuim), compound 2 - $(F-PMA)_2PbI_4$ ($F-PMA^+$ – bis(4-fluorophenylmethylammonium), compound 3 - $GUACsPbI_4$ ($GUA^+$ – guanidinium) and compound 4 - $MAPbI_3$ ($MA^+$ – methylammonium). A detailed information about the synthesis is in the supporting information file. The structure types and crystal structures of the synthesized compounds are not new and were previously refined from single crystals in [34–37]. Laboratory data were collected on a D8 Bruker diffractometer with the CuKα radiation in the 3–30° 2θ range. The diffraction patterns of the compounds are shown in the figure S2 in SI. The unit cell parameters of each compound were obtained using the Jana program [38]. (hkl) reflexes were indicated by comparison with the previously published literature data. The unit cell parameters and the structural features of each compound are reported in Supporting information file.

*Comparison of the decision tree model with other classification algorithms*

In order to compare the decision tree (DT) approach with alternative machine learning (ML) models, a Random Forest (RF) [39], Extra Trees [40], XGB [41], and Cat Boost [42] models were constructed and exhibited comparable accuracies to the DT classifier (see Table S1). When classifying theoretical XRD patterns into three classes based on the dimensions of the inorganic substructure, all five ML classification algorithms demonstrated analogous performance in terms of quality metrics such as accuracy, balanced accuracy, precision, recall, F1 score, and Matthews correlation coefficient (MCC) (Figure 8). Evaluation of the algorithms on experimental XRD patterns revealed that the DT, Random Forest, and Cat Boost classifiers achieved superior results (11 true positive outcomes out of 11) compared to the Extra Trees and XGB algorithms (8/11 and 10/11) (see Table S1). Furthermore, the accuracy of determining the hybrid halide structure type was consistent across all algorithms in terms of quality metrics (see Figure 9). Notably, the algorithms displayed equal sensitivity to variations in the number of crystal structures within a class. Therefore, enhancing the performance of ML models necessitates continuous augmentation of the dataset with newly encountered structure types that are currently underrepresented.

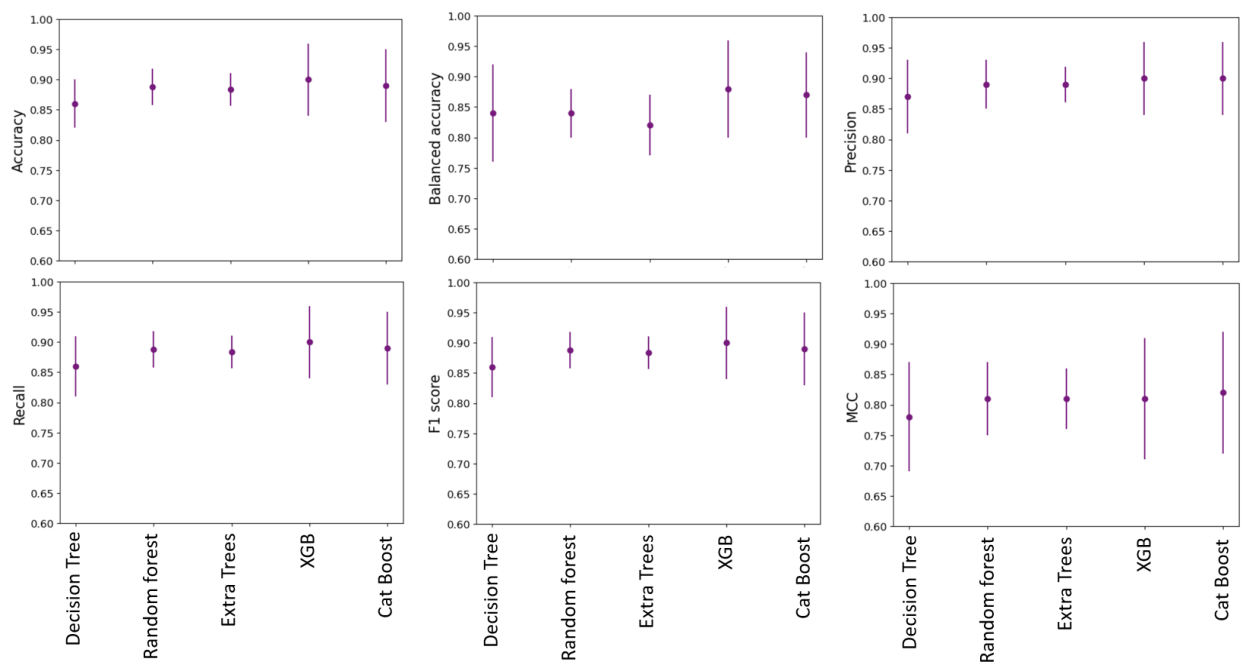

Figure 8. Comparison of model performance of four ML algorithms of classifications XRD patterns by dimensionality on three classes (1D, 2D and 3D). Vertical lines indicate error bars.

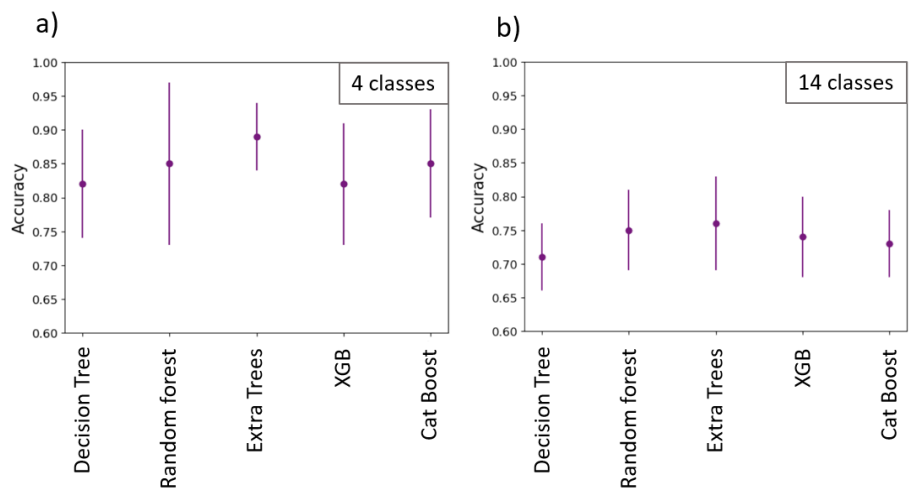

Figure 9. Comparison of model performance of four ML algorithms of classifications XRD patterns by structure type on 4 (a) and 14 classes (b). Vertical lines indicate error bars.

## Conclusions

In this work, we developed a new classification ML algorithm (decision tree) to quickly identify the diffraction patterns of hybrid lead halides by the most common structure types. The average accuracy of our ML algorithm in predicting the dimensionality of inorganic substructure, type of connection of lead halide and inorganic substructure topology by theoretically calculated XRD pattern among 14 most common structure types reaches 0.86±0.05, 0.827±0.028 and 0.71±0.05, respectively. When the training/testing set size is constrained by specific structural types and the DT model is trained on the four predominant structural types with more than five structures each, alongside a test sample comprising more than two crystal structures of identical topological types, the algorithm demonstrates an average accuracy of 0.82±0.08 at discriminating inorganic substructure topology among the four structure types that dominate the dataset. Moreover, our decision tree classification model reached accuracies of 1.00 and 0.82 for dimension and structure type prediction using experimental XRD from powder and thin films with distinct texturing, respectively. Thus, our approach can significantly simplify and accelerate the interpretation of XRD data for hybrid lead halides.

## Acknowledgements

This work was supported by the Russian Science Foundation (grant № 23-73-01212, https://rscf.ru/project/23-73-01212/). XRD studies were performed using the equipment of the Joint Research Centre for Physical Methods of Research of the Kurnakov Institute of General and Inorganic Chemistry of the Russian Academy of Sciences (JRC PMR IGIC RAS).

## Conflict of Interest

The authors declare no conflict of interest.

**Machine Learning Recognition of hybrid lead halide perovskites and perovskite-related structures out of X-ray diffraction patterns**

Marchenko E.I.[a,b], Korolev V.V.[c], Kobeleva E. A.[b], Belich N.A.[a], Udalova N.N.[a], Eremin N.N.[b,e], Goodilin E.A.[a,d], Tarasov A.B.[a,d]*

[a] *Laboratory of New Materials for Solar Energetics, Department of Materials Science, Lomonosov Moscow State University; 1 Lenin Hills, 119991, Moscow, Russia;*
[b] *Department of Geology, Lomonosov Moscow State University; 1 Lenin Hills, 119991, Moscow, Russia*
[c] *MSU Institute for Artificial Intelligence, Lomonosov Moscow State University; 119192, Moscow, Russia*
[d] *Department of Chemistry, Lomonosov Moscow State University; 1 Lenin Hills, 119991, Moscow, Russia*
[e] *Institute of Geology of Ore Deposits, Petrography, Mineralogy, and Geochemistry, Russian Academy of Science, Moscow, Russia*
*e-mail: alexey.bor.tarasov@yandex.ru*

# Supporting Information

*Evaluation of machine learning models*

The evaluation of machine learning algorithms plays a crucial role in assessing their performance. However, it can be challenging, especially in scenarios where limited or no access to real-world data exists. In such cases, additional human effort is required to assess the model performance. In classification tasks, the evaluation is typically carried out by splitting the dataset into a training set and a test set. The machine learning algorithm is trained on the training set, while the test set is used to calculate performance indicators that assess the performance of the model. One common challenge faced by machine learning algorithms is the availability of limited training and test data. This can impact the algorithm's generalization capabilities and lead to overfitting or underfitting. It's crucial to have enough data for training and testing to have a good machine-learning model. Evaluating the performance of a machine learning model involves considering multiple factors. While there is no perfect indicator applicable to every scenario, several important factors are considered. These factors include:

• Accuracy: Measures the proportion of correctly classified instances, providing
an overall assessment of the algorithm's performance.
• Precision: Evaluates the algorithm's ability to correctly identify positive instances within a given class, indicating its effectiveness in minimizing false positives.
• Recall: Assesses the algorithm's ability to identify all positive instances within
a given class, indicating its effectiveness in minimizing false negatives.
• F1 score: Combines precision and recall into a single metric, providing a balanced measure of a classifier's performance

It is important to select appropriate evaluation metrics based on the specific problem domain and objectives of the machine learning algorithm. The choice of evaluation metrics should align with the desired outcomes and provide meaningful insights into the algorithm's performance. The performance of the machine learning model, could be compromised due to the issues such as class imbalance, for example, if the data set is dominated by the class-2D over the classes 0D and 1D. In this case, it is more advantageous to have a model that is able to predict the positive instances for each of those classes with high accuracy rather than using a metric that assesses overall predictions. Accuracy measures the overall correctness of the predictions, which is not a suitable performance metric when class distribution is imbalanced. In recall the score is calculated on the true positives and false negatives values. False negatives describe the ability of the model to identify the positive instances correctly. Recall is not crucial for this type of classification but may be important for certain scenarios such as medical diagnosis, where false negative class is more important. In this classification model, the focus is on achieving accurate predictions for each class that best match the definitions of precision where incorrectly predicted positive instance gives a significant reduction in the model performance. Therefore, precision over recall and accuracy were selected as performance metrics.

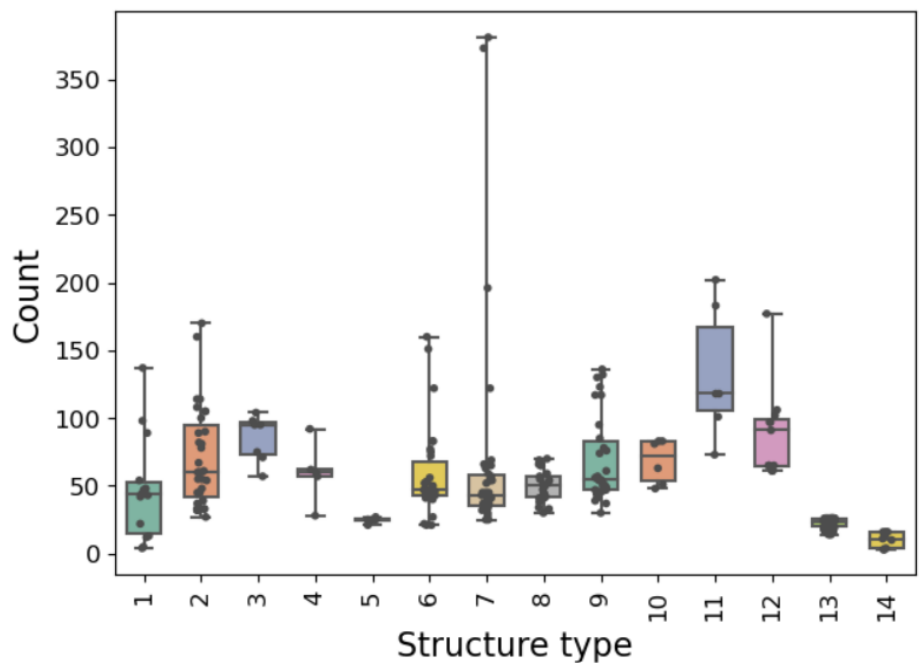

**Figure S1.** Boxplot of the number of reflections on the simulated XRD pattern in the angle range 3-30 ° 2θ. The x axis shows the different structure types according to Table 1.

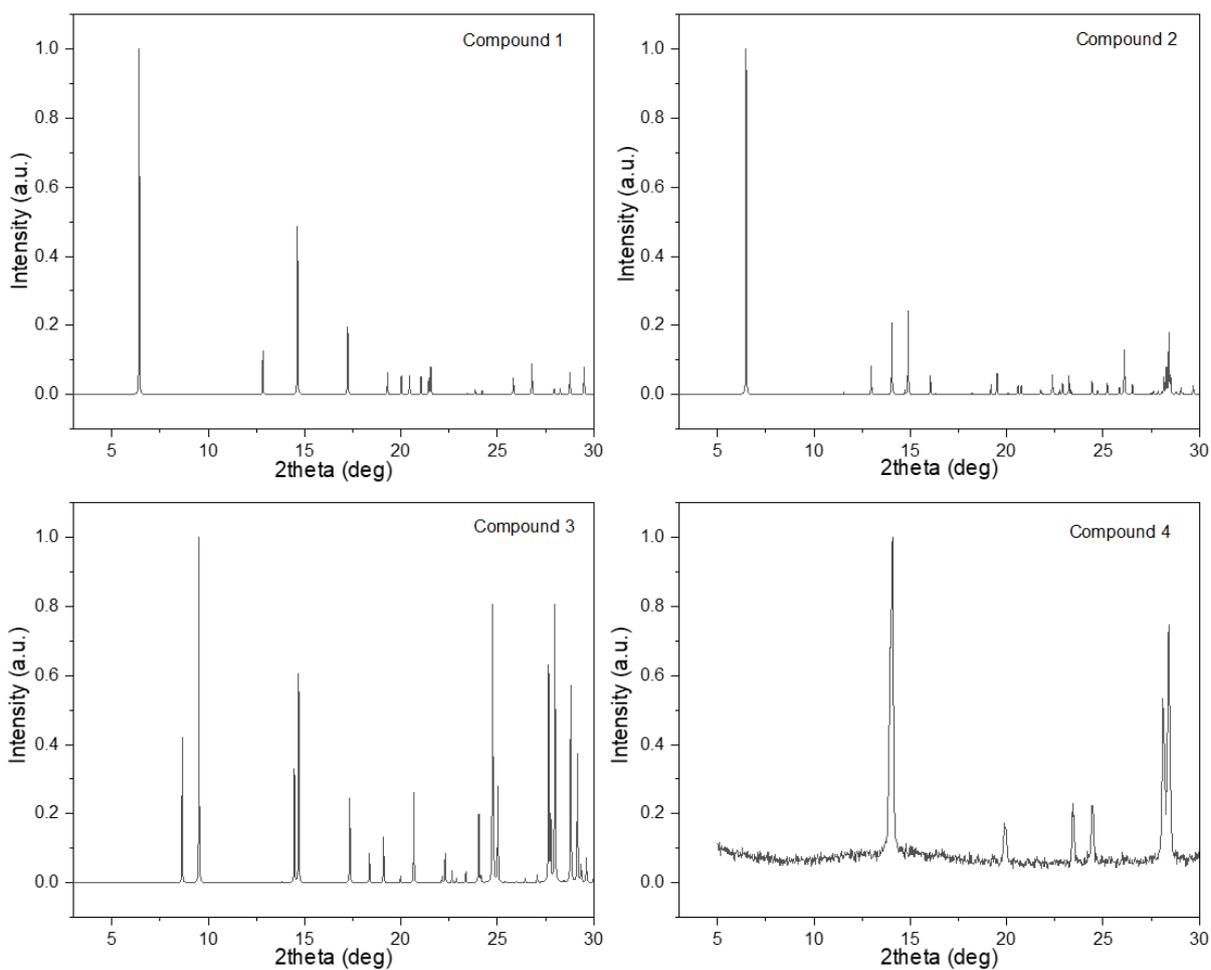

**Figure S2.** Experimental XRD patterns of compounds 1-4 (Table 2) from thin films at Cu Kα radiation using for testing of ML algorithms.

## Synthesis of hybrid lead halide thin films

Compound 1: 1M solution of BAI and PbI$_2$ in 2:1 molar ratio was spin-coated at 6000 rpm for 20s without antisolvent. Film was annealed at 100 °C for 10 min.

Compound 2: 1M solution of (F-PMA)I and PbI$_2$ in 2:1 molar ratio was spin-coated at 6000 rpm for 20s without antisolvent. Film was annealed at 100 °C for 10 min.

Compound 3: 1M solution of GUAI, CsI and PbI$_2$ in 1:1:1 molar ratio was spin-coated at 6000 rpm for 20s without antisolvent. Film was annealed at 100 °C for 10 min.

Compound 4: 1.5M solution of MAI and PbI2 in equimolar ratio in DMF/DMSO (4:1 v/v) was spin-coated onto previously cleaned glass substrates at 6000 rpm with quick addition of 100 µl chlorobenzene antisolvent at 10th second of spinning. Film then was annealed at 100°C for 30 min.

## The crystal structures determined by X-ray diffraction

The unit cell parameters and space groups were determined for synthesized compounds. The data obtained are consistent with the literature data on the structure refining of these compounds.

The crystal structures of these compounds belong to the 4 most common structure types: 2D inorganic substructure of (100) type with n=1 and (110) corrugated 2x2 motifs, 1D inorganic substructure with chains of octahedra connected along vertices and 3D crystal structures with perovskite structure type (Figure S3).

**Compound 1.** Composition – [CH$_3$(CH$_2$)$_3$NH$_3$]$_2$PbI$_4$ (BA$_2$PbI$_4$). The crystal structure was refined using the ordered model previously published by [1] in the *Pbca* space group. The refinement of the collected powder pattern led to lattice constants with *a* = 8.8632 Å, *b* = 8.6814 Å, *c* = 27.5692 Å and R$_p$ = 2.56, R$_{wp}$ = 8.28, GOF = 0.05.

**Compound 2.** Composition - [FC$_6$H$_5$CH$_2$NH$_3$]$_2$PbI$_4$ (F-PMA)$_2$PbI$_4$). The crystal structure was refined using the ordered model previously published by [2] in the *P2$_1$/c* space group. The refinement of the collected powder pattern led to lattice constants with *a* = 8.6973 Å, *b* = 9.2461 Å, *c* = 27.5309 Å, β=97.603 ° and R$_p$ = 3.11, R$_{wp}$ = 11.15, GOF = 0.01.

**Compound 3.** Composition - Cs[C(NH$_2$)$_3$]PbI$_4$ CsGUAPbI$_4$). The crystal structure was refined using the ordered model previously published by [3] in the *Pnnm* space group. The refinement of the collected powder pattern led to lattice constants with *a* = 12.7425 Å, *b* = 18.6066 Å, *c* = 12.1767 Å and R$_p$ = 3.29, R$_{wp}$ = 12.48, GOF = 0.15.

**Compound 4.** MAPbI$_3$ (MA$^+$ – methylammonium). The crystal structure was refined previously using the disordered model previously published by [4] in the *P4/mcm* space group with cell parameters *a* = 8.85728 Å and *c* = 12.65104 Å. The XRD pattern of compound 4 is in good agreement with the model proposed in the [37] (Figure S3).

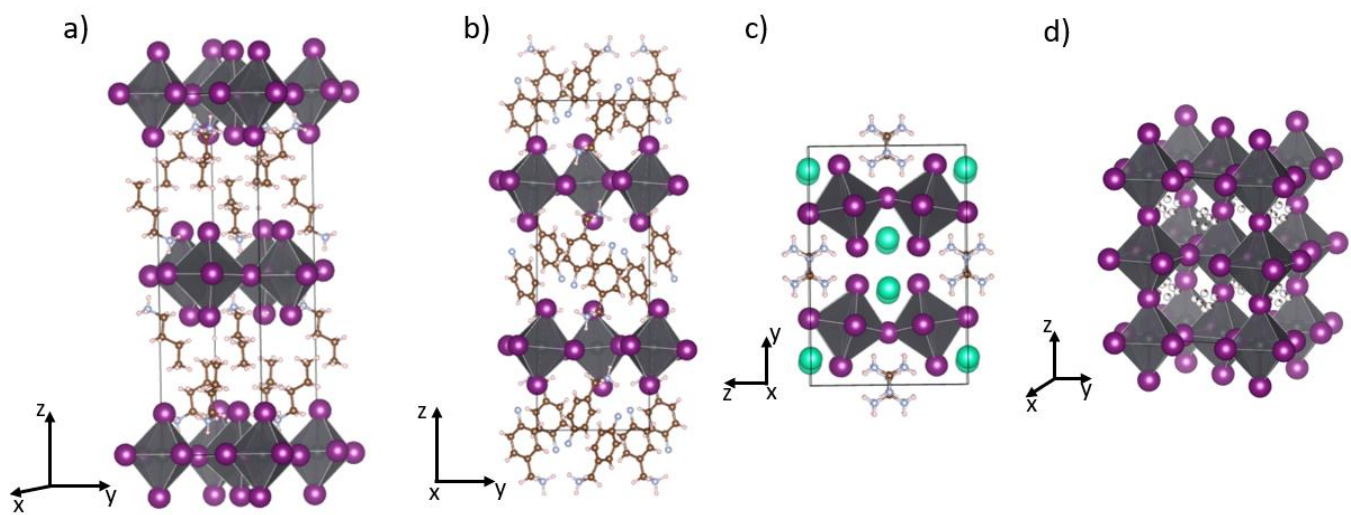

**Figure S3.** The crystal structures of synthesized compounds: a) BA$_2$PbI$_4$ – 2D (100), b) (F-PMA)$_2$PbI$_4$ – 2D (100), c) CsGUAPbI$_4$ 2D - (100), d) MAPbI$_3$ – 3D.

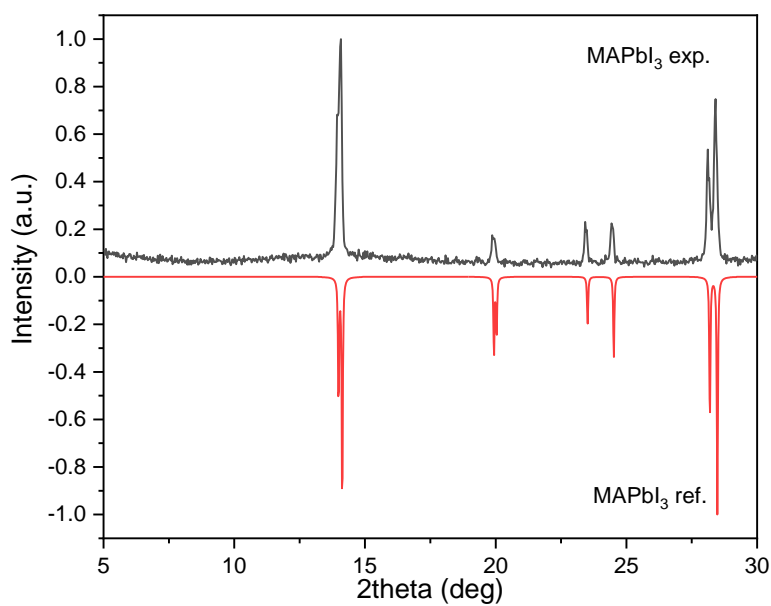

**Figure S4.** Comparison of MAPbI$_3$ experimental XRD pattern from this work (black) and the XRD pattern of model proposed in [37] (red).

**Table S1.** Comparison of the performance of different ML classification algorithms.

**Dimensionality (3 classes)**

|  | accuracy | balanced accuracy | precision | recall | F1 score | MCC | accuracy (exp. data) |
|---|---|---|---|---|---|---|---|
| DecisionTreeClassifier | 0.86±0.05 | 0.84±0.08 | 0.87±0.06 | 0.86±0.05 | 0.86±0.06 | 0.78±0.09 | 11/11 |
| RandomForestClassifier | 0.888±0.03 | 0.84±0.04 | 0.89±0.04 | 0.888±0.030 | 0.882±0.031 | 0.81±0.06 | 11/11 |
| ExtraTreesClassifier | 0.884±0.027 | 0.82±0.05 | 0.890±0.029 | 0.884±0.027 | 0.871±0.035 | 0.81±0.05 | 8/11 |
| XGBClassifier | 0.90±0.06 | 0.88±0.08 | 0.90±0.06 | 0.90±0.06 | 0.89±0.06 | 0.83±0.10 | 10/11 |
| CatBoostClassifier | 0.89±0.06 | 0.87±0.07 | 0.90±0.06 | 0.89±0.06 | 0.89±0.06 | 0.82±0.10 | 11/11 |

**Dimensionality (4 classes)**

|  | accuracy | balanced accuracy | precision | recall | F1 score | MCC |
|---|---|---|---|---|---|---|
| DecisionTreeClassifier | 0.76±0.07 | 0.67±0.13 | 0.81±0.06 | 0.76±0.07 | 0.78±0.07 | 0.65±0.10 |
| RandomForestClassifier | 0.85±0.06 | 0.66±0.09 | 0.84±0.07 | 0.85±0.06 | 0.83±0.08 | 0.76±0.10 |
| ExtraTreesClassifier | 0.846±0.033 | 0.65±0.06 | 0.84±0.05 | 0.846±0.033 | 0.83±0.05 | 0.76±0.05 |
| XGBClassifier | 0.85±0.05 | 0.69±0.06 | 0.85±0.05 | 0.85±0.05 | 0.84±0.05 | 0.76±0.07 |
| CatBoostClassifier | 0.86±0.06 | 0.70±0.09 | 0.87±0.06 | 0.86±0.06 | 0.86±0.05 | 0.78±0.09 |

**Type of structure (14 classes)**

|  | accuracy | balanced accuracy | precision | recall | F1 score | MCC |
|---|---|---|---|---|---|---|
| DecisionTreeClassifier | 0.71±0.05 | 0.66±0.06 | 0.723±0.031 | 0.71±0.05 | 0.696±0.034 | 0.68±0.05 |
| RandomForestClassifier | 0.75±0.06 | 0.66±0.09 | 0.74±0.05 | 0.75±0.06 | 0.73±0.05 | 0.72±0.07 |
| ExtraTreesClassifier | 0.76±0.07 | 0.69±0.09 | 0.74±0.06 | 0.76±0.07 | 0.74±0.06 | 0.73±0.08 |
| XGBClassifier | 0.74±0.06 | 0.68±0.10 | 0.74±0.05 | 0.74±0.06 | 0.72±0.06 | 0.71±0.07 |
| CatBoostClassifier | 0.73±0.05 | 0.67±0.07 | 0.746±0.014 | 0.73±0.05 | 0.72±0.04 | 0.70±0.06 |

**Type of structure (4 classes)**

| | accuracy | balanced accuracy | precision | recall | F1 score | MCC | accuracy (exp. data) |
|---|---|---|---|---|---|---|---|
| DecisionTreeClassifier | 0.82±0.08 | 0.83±0.11 | 0.86±0.10 | 0.82±0.08 | 0.81±0.10 | 0.75±0.12 | 9/11 |
| RandomForestClassifier | 0.85±0.12 | 0.85±0.15 | 0.87±0.13 | 0.85±0.12 | 0.84±0.14 | 0.79±0.16 | 8/11 |
| ExtraTreesClassifier | 0.89±0.05 | 0.83±0.11 | 0.89±0.06 | 0.89±0.05 | 0.88±0.05 | 0.84±0.08 | 9/11 |
| XGBClassifier | 0.82±0.09 | 0.78±0.11 | 0.81±0.10 | 0.82±0.09 | 0.79±0.11 | 0.75±0.13 | 9/11 |
| CatBoostClassifier | 0.85±0.08 | 0.85±0.11 | 0.87±0.08 | 0.85±0.08 | 0.83±0.10 | 0.79±0.11 | 7/11 |

**Connection of octahedra (6 classes)**

| | accuracy | balanced accuracy | precision | recall | F1 score | MCC |
|---|---|---|---|---|---|---|
| DecisionTreeClassifier | 0.827±0.028 | 0.67±0.12 | 0.82±0.05 | 0.827±0.028 | 0.82±0.04 | 0.72±0.05 |
| RandomForestClassifier | 0.82±0.04 | 0.57±0.12 | 0.78±0.06 | 0.82±0.04 | 0.79±0.05 | 0.70±0.07 |
| ExtraTreesClassifier | 0.84±0.04 | 0.65±0.09 | 0.81±0.05 | 0.84±0.04 | 0.81±0.04 | 0.73±0.07 |
| XGBClassifier | 0.842±0.029 | 0.64±0.11 | 0.83±0.04 | 0.842±0.029 | 0.828±0.031 | 0.74±0.05 |
| CatBoostClassifier | 0.85±0.06 | 0.72±0.16 | 0.86±0.07 | 0.85±0.06 | 0.84±0.07 | 0.76±0.10 |